\documentstyle[aps,preprint,floats,psfig]{revtex}
\begin{document}
\newcommand {\be}{\begin{equation}}
\newcommand {\ee}{\end{equation}}
\newcommand {\bea}{\begin{eqnarray}}
\newcommand {\eea}{\end{eqnarray}}
\newcommand {\nn}{\nonumber}

\draft
%
%
%
%

\title{Dynamical Properties of Spin-Orbital Chains in a Magnetic Field}

\author{Weiqiang Yu and Stephan Haas}
\address{Department of Physics and Astronomy, University of Southern
California, Los Angeles, CA 90089-0484
}

\date{\today}
\maketitle

\begin{abstract}
The excitation spectrum of the one-dimensional spin-orbital model 
in a magnetic field is studied, using a recently developed dynamical 
density matrix renormalization group technique. The method is employed 
on chains with up to 80 sites, and examined for test cases such as the 
spin-1/2 antiferromagnetic Heisenberg chain, where the excitation 
spectrum is known exactly from the Bethe Ansatz. In the spin-orbital
chain, the characteristic dynamical response depends strongly on the 
model parameters and the applied magnetic field. The coupling between 
the spin and orbital degrees of freedom is found to influence the 
incommensuration at finite magnetizations. In the regions of the phase 
diagram with only massive spin and orbital excitations, a finite field 
is required to overcome the spin gap. An incommensurate orbital mode is 
found to become massless in this partially spin-polarized regime, 
indicating a strong coupling between the two degrees of freedom.
In the critical region with three elementary gapless excitations, a
prominent particle-hole excitation is observed at higher energies,
promoted by the biquadratic term in the model Hamiltonian of the
spin-orbital chain.

\end{abstract}

\section{Introduction}

There is a wide variety of quasi-one-dimensional spin 
systems with short-ranged and quasi-long-ranged quantum liquid states.
Examples for short-ranged compounds include the spin-Peierls chain, 
even-leg spin-1/2 Heisenberg ladders, and integer-spin Heisenberg chains.
In contrast, half-odd-integer-spin chains and odd-leg spin-1/2 Heisenberg 
ladders are known to have a quantum critical ground state with
quasi-long-range-ordered spin correlations. Many quasi-one-dimensional 
models, such as the antiferromagnetic spin-1/2 Heisenberg chain with 
frustrating longer-range exchange interactions, have a phase diagram 
which contains both types of disordered ground state, depending on the 
parameters. Particular recent attention has focussed on spin-1/2 
Heisenberg chains coupled to orbital degrees of freedom.
\cite{pati,yamashita,itoi,azaria,azaria2,yamashita2,frischmuth,orignac,lee} 
These systems are known to have a rich phase diagram, containing 
short-ranged and quasi-long-ranged regions, including an SU(4)-symmetric 
integrable point at a particular combination of the exchange parameters. 

In this work, we study the spin and orbital excitation spectra of 
spin-orbital chains in a magnetic field. By applying a newly developed 
dynamical density matrix renormalization group (DMRG) technique,
\cite{hallberg,kuhner} it is found
that the admixture of the orbital 
degrees of freedom has a profound influence on both the
field-dependent incommensuration, 
and the dispersion and spectral weight
of the spin triplet modes, although
only the spins couple directly to the applied magnetic field. 
Furthermore, this study 
is intended to provide useful information for inelastic neutron scattering
studies on orbitally degenerate 
materials such as $\rm Na_2Ti_2Sb_2O$\cite{axtell}, where
the d-electron of the Ti$^{3+}$ ion occupies an $e_g$ doublet, and 
$\rm NaV_2O_5$, where orbital degeneracy appears to arise from the coupling of 
two types of $\rm VO_5$ chains.\cite{pati,isobe,fujii,smolinski,augier}
As long as experimentally available magnetic fields ($\sim$ 30 Tesla) are
sufficiently large to overcome the respective spin gaps,
our results for the dynamical properties of the incommensurate phases
in the one-dimensional spin-orbital model may be tested on these or 
suitably related compounds. 

In this paper, the Hamiltonian of the one-dimensional
spin-orbital model in a magnetic field is studied. This can be written as
\bea 
H = J_1 \sum_{i=1}^N {\bf S}_i \cdot {\bf S}_{i+1}
  + J_2 \sum_{i=1}^N {\bf T}_i \cdot {\bf T}_{i+1} 
  + K \sum_{i=1}^N \left( {\bf S}_i \cdot {\bf S}_{i+1} \right)
\left( {\bf T}_i \cdot {\bf T}_{i+1} \right)
  - h \sum_{i=1}^N S^z_i,
\eea
where the spin-1/2 operators ${\bf S}_i$ and the pseudo-spin-1/2 operators
${\bf T}_i$ describe the 
spins and orbits at a site $i$.\cite{kugel}
The first two terms are the Hamiltonians 
for (pseudo-)spin-1/2 Heisenberg chains with exchange constants $J_1$ for 
the spins and $J_2$ for the orbitals. The third term couples these two sets 
of degrees of freedom with a constant $K$. The fourth term is the Zeeman 
coupling of the spins to an external magnetic field aligned with the 
$z$-direction. In the absence of a magnetic field, this Hamiltonian has 
an SU(4) symmetry for the particular parameter combination $J_1 = J_2 = K/4$.
This symmetry reduces to SU(2)$\times$U(1) in an intermediate regime 
($ 0 < h < h_c$), and to SU(2) at very high magnetic fields where the 
spins are completely aligned. If the parameter set ($J_1,J_2,K$) is 
chosen away from the SU(4) point, the Hamiltonian remains 
invariant under SU(2)$\times$SU(2). Furthermore, there is a
$Z_2$ symmetry about $J_1 = J_2$ for vanishing magnetic fields.

Recent studies have concentrated on the phase diagram\cite{pati,yamashita,itoi}
and low-energy excitations\cite{yamashita2,li} of $H$. At the integrable 
SU(4) point, the spectrum of excitations has been obtained
from the Bethe Ansatz equations\cite{li}. Effective low-energy theories
were obtained from bosonization\cite{azaria,azaria2}, and 
approximate solutions from mean-field theory\cite{brink}.
In addition, a variety of numerical techniques,
including series expansions\cite{pati}, numerical 
diagonalization\cite{yamashita2}, 
Monte Carlo\cite{frischmuth}, and density matrix renormalization 
group\cite{yamashita2}, has been employed to analyze the ground state 
and low-energy properties of this Hamiltonian. 

In this work, a recently developed density matrix renormalization group 
technique is applied to study the full spin and orbital
dynamics of the one-dimensional spin-orbital 
model in a magnetic field for arbitrary parameters ($J_1,J_2,K,h$). The 
advantages of this method are that (i) relatively large Hilbert spaces
can be studied (typical system size $\approx$ 80 sites), 
(ii) the full excitation spectra can be obtained, 
and (iii) there are no constraints concerning low energies, small 
deviations from the SU(4) point, or the magnitude of the applied
magnetic field. However, the spectra calculated with this method do 
exhibit some finite-size effects, as expected for low-dimensional
quantum systems. Here we concentrate on the characteristic features 
of the dynamical spin and orbital structure factors for the various 
regimes of $H$. These quantities are defined as
\bea
S(k,\omega ) = - \frac{1}{\pi} \lim_{\epsilon \rightarrow 0}
{\rm Im}  \langle \Psi_0 | S^z_{-k} \frac{1}{\omega + i \epsilon - H + E_0}  
S^z_k | \Psi_0 \rangle,\\
T(k,\omega ) = - \frac{1}{\pi} \lim_{\epsilon \rightarrow 0}
{\rm Im} \langle \Psi_0 | T^z_{-k} \frac{1}{\omega + i \epsilon - H + E_0}  
T^z_k | \Psi_0 \rangle,
\eea
where $|\Psi_0 \rangle$ is the ground state, $E_0$ its energy, $S^z_k$ and
$T^z_k$ are the Fourier transforms of the local spin and orbit
operators $S^z_i$ and $T^z_i$, and a small but finite broadening 
factor $\epsilon = 0.05 J_1$ is chosen to represent the poles in 
$S^z_k$ and $T^z_k$.

In general, for the properties of the orbitally degenerate Hamiltonian
$H$ to be applicable to candidate inorganic materials
such as $\rm Na_2Ti_2Sb_2O$ and $\rm NaV_2O_5$,
the orbital energy levels must be nearly degenerate on the 
scale of the spin dynamics, which have characteristic energies 
in the range of $J_1 \approx$ 10-100 meV.
While external magnetic fields couple directly to the 
spin degrees of freedom, applied external pressure
could be a ``dual" perturbation,
primarily affecting the orbitals. It would
therefore be of great interest to explore the dynamical
inelastic neutron scattering
response in candidate spin-orbital compounds subject to
a magnetic field as well as to external pressure. 
 
In the following section, we discuss briefly the numerical renormalization
group method which is used to obtain the spin dynamics of the
spin-orbital chain in a magnetic field. In section III, the field-dependent
spectra are studied in the different parameter regimes ($J_1,J_2,K,h$). 
A summary and conclusions are given in section IV.

\section{Calculation of Dynamical Properties with the Density Matrix
Renormalization Group Technique}

The density matrix renormalization group (DMRG) method is known
to be a precise numerical technique to calculate the ground state
properties of quasi-one-dimensional systems such as chains and
ladders.\cite{white}
Recently, Hallberg\cite{hallberg} and K\"uhner and White
\cite{kuhner} have proposed extensions of this method to extract
dynamical correlation functions, based on continued-fraction 
expansions of the dynamical Green function. In this paper, we use a 
variant of this technique to calculate the momentum and 
frequency-dependent spectrum of the spin-orbital model in one 
spatial dimension. 

The zero-temperature, time-dependent correlation function of 
a momentum-dependent operator $\hat{A}_k$ is defined as
\bea
A(k,t - t') \equiv \langle \Psi_0 | \hat{A}^{\dagger}_{-k}(t) \hat{A}_k(t')
| \Psi_0 \rangle ,
\eea
where $| \Psi_0 \rangle$ is the ground state. 
In the following, $\hat{A}_k$ corresponds either to
the spin operator $S^z_k$ or to the orbital
operator $T^z_k$ . The Fourier transform of this correlation 
function can be expanded in the form of a continued fraction,
\bea
A(k,\omega ) = - \frac{1}{\pi} {\rm Im}
\frac{\langle \Psi_0 | \hat{A}^{\dagger}_{-k} \hat{A}_k | \Psi_0 \rangle }
{\omega + i\epsilon - a_0 - \frac{b_1^2}{\omega + i\epsilon - a_1 - ...}},
\eea
where $\epsilon \rightarrow 0$ is a small broadening factor. The 
coefficients $a_n$ and $b_n$ are then obtained from the recursion 
relation
\bea
| f_{n+1} \rangle = H | f_n \rangle - a_n | f_n \rangle
- b_n^2 | f_{n-1} \rangle ,
\eea
with $| f_0 \rangle \equiv \hat{A}_k | \Psi_0 \rangle$ and
$b_0 \equiv 0$. This procedure is very similar to numerical diagonalization
calculations. However, there are specific technical problems intrinsic to  
the DMRG method which need to be addressed.\cite{hallberg,kuhner}

One complication arises when open boundary conditions (OBC) are used, 
which is usual in DMRG calculations to yield the highest possible
numerical precision. In this case, 
the momentum is not a well-defined quantum number, and each 
momentum-dependent operator $\hat{A}_k$ generates 
wave packets of finite width, centered around $k$. 
To remedy this problem, Hallberg has used periodic boundary conditions (PBC)
to obtain a basis with a good momentum quantum number, whereas
K\"uhner and White have used OBCs and the
Parzen filter to eliminate the artificial surface
excitations on the surface boundaries. In this paper, 
we adapt OBCs combined with the Parzen filter
to obtain the full dynamical spectra associated with $S^z_k$ and $T^z_k$.
OBCs are known to give more precise
ground states than PBCs, and therefore larger 
clusters can be studied, while the width of the corresponding wave
packets is greatly reduced. 

Further improvement of the dynamical DMRG method may be obtained by 
using more target states besides the ground state, implemented with 
``correction vectors''.\cite{kuhner,pati2} We find that a rather large cut-off 
dimension of the density matrix is needed to reach convergence in the 
spin-orbital chain for the target states at higher energy. These convergence 
problems tend to be exacerbated for excitations away from the dominant 
wave vectors. However, with the algorithm used in this work, the full 
structure of the excitation spectrum can be obtained with rather high 
accuracy, even if only the ground state is kept as a target state, 
although the relative overlap (matrix elements) with higher-energy 
states is reduced. 

We have tested the algorithm on the spin-1/2 antiferromagnetic 
Heisenberg chain in a magnetic field. The Hamiltonian for this system 
corresponds to the parameter choice ($J_1=J,J_2=K=0$) for the spin-orbital 
model. One observes that the spin-triplet excitation spectrum of this 
simple model matches very well the exact results from the Bethe 
Ansatz.\cite{cloiseaux} A more detailed discussion of this spectrum will be
presented in the next section. Chains with $N = 80$ spins were studied. The 
cut-off factor, {\it i.e.} the number of states kept per block, was taken 
to be $m = 160$. We have also used the Lanczos method on lattices of up to 
16 sites to check the accuracy of the DMRG results
for other non-trivial parameter sets of the spin-orbital chain.

\section{Dynamical Spin Structure Factor of the Spin-Orbital Chain}

The rich phase diagram of the one-dimensional spin-orbital model, 
represented by the Hamiltonian $H$ in Eq. 1, has been the subject 
of a number of recent studies.\cite{pati,yamashita,itoi} 
The main result of these works is summarized in Fig. 1, which we have
adapted from Ref.\cite{itoi}. The various phases can be characterized
by spin and orbital quantum numbers $(S,S^z)$ and $(T,T^z)$, and
by effective Luttinger parameters in each channel, such as gaps 
$(\Delta_S, \Delta_O)$, soliton
velocities $(v_S,v_O)$, and radii of compactification.
For finite values of the spin-orbital interaction $K$, the 
spin and orbital degrees of freedom are coupled, and the excitation 
spectra generally contain strongly admixed spin and orbital modes.

Let us first briefly describe the phase diagram of $H$.
Phase I is ``ferromagnetic" in the spin and orbital channels,
{\it i.e.} both degrees of freedom are 
fully polarized, and there is an infinite degeneracy in the ground state.
In phase II, the spin degrees of freedom remain polarized,
whereas the orbitals are in a quasi-long-range-ordered antiferromagnetic
state with a characteristic
orbiton velocity $v_O$. In phase III, the 
roles of spins and orbitals are interchanged. 
Phase IV is fully gapped in all channels ($\Delta_S \neq 0, \Delta_O \neq 0$) 
because of spontaneous dimerization. In particular, this phase contains
a Majumdar-Ghosh point at $J_1=J_2=3K/4$ which has
a matrix-product ground state.
\cite{mg,kolezhuk} It has recently been shown
\cite{yamashita,itoi} that phase V contains 3 gapless excitations with
non-universal, parameter-dependent soliton velocities. At the SU(4)-symmetric
point (filled circle in Fig. 1), these three modes have one common velocity,
$v = \pi J_1/2 = \pi J_2/2  = \pi K/8$.\cite{sutherland} 

Let us now turn to the excitation spectra of $H$ in these regimes, 
starting with phase III where the orbital degrees of freedom are fully
polarized. In Fig. 2, the spectrum
traced by the dynamical spin structure factor is shown
on a lattice of 80 sites, using the parameter set ($J_1=-J_2=K/2$). 
The poles have been given a width
of $\epsilon = 0.05 J_1$, and the color scheme highlights the areas 
of high spectral intensity with bright colors. Since in this phase,
the orbital degrees of freedom couple only trivially with the spins,
the analysis turns out to be particularly simple. The spin excitation 
spectrum of this system corresponds to a spin-1/2, antiferromagnetic 
Heisenberg chain with exchange coupling $J_1$, which 
is dominated by a low-energy  2-spinon continuum (see Fig. 
2(a)).\cite{bougourzi}
The elementary excitations of this system are
spin-1/2 solitons with dispersion $\omega_{\sigma} (q) =
3 \pi J_1 \sin{(q)} /4$ for $q \in [0,\pi ]$.
Thus the 2-spinon continuum has a lower cut-off 
at $\omega_l (k) = 3 \pi J_1 |\sin{(k)}|/4$ and an upper cut-off at
$\omega_u (k) = 3 \pi J_1 |\sin{(k/2)}| /2$.\cite{cloiseaux,haas}
The dominant singularity of the dynamical spin structure factor 
at wave vector $\pi$ diverges according to $S(\pi,\omega ) \propto 
\sqrt{\ln{(\omega )}}/\omega $ as $\omega \rightarrow 0$.

In a finite magnetic field, this mode becomes incommensurate with 
nesting vectors $2k_F = \pi ( 1 \pm 2 m)$, where $m \in [0, 1/2]$ 
is the magnetization. This behavior is confirmed by our dynamical 
DMRG calculation, as shown in Fig. 2(b). Here, a finite magnetization, 
$m = 4/40$, leads to an incommensuration with $2k_F = 12 \pi/10$ and 
$8 \pi /10$. Because of the relatively large system size ($N = 80$),
the spectra appear to be almost continuous, as expected in the 
thermodynamic limit. However, finite-size effects are still visible, 
causing gaps and uneven spectral weight distributions at higher energies. 
Furthermore, the momentum is quantized in 80 slices of size $\pi/40$. 
In spite of these limitations,
the agreement of the numerical results with the exactly known spectra is 
encouraging, and gives confidence to proceed to parameter 
regimes with more complex excitation spectra.

At the SU(4) point 
($J_1 = J_2 = K/4$),  the spectrum is also known exactly from the Bethe
Ansatz. A comparison of the spectra obtained
using the DMRG method with the exact results 
can thus help to further evaluate the accuracy of the
numerical approach. In Fig. 3, the spectrum
traced by the dynamical spin structure factor is shown
on a lattice of 80 sites.
From the solution of the Bethe Ansatz equations, it is known that
the elementary excitations of the SU(4) chain are 4-fold degenerate 
spin-1/2 spinons ($\sigma$) and pseudo-spin-1/2 orbitons ($\tau$), and  
6-fold degenerate spin-1 or orbital-1 solitons ($\nu$).\cite{li}. The 
corresponding dispersions of these three elementary excitations are
\bea
\omega_{\sigma} (q) &=&  \frac{J \pi}{2}  \left[ \sqrt{2}
\cos{(-q + 3\pi/4)} + 1\right], q \in [0, 3\pi/2]  \nn \\
\omega_{\tau} (q) &=&  \frac{J \pi}{2}  \left[ \sqrt{2}
\cos{(-q + \pi/4)} -1\right], q \in [0, \pi/2]   \\
\omega_{\nu} (q) &=&  \frac{J \pi}{2} \cos{(- q + \pi/2)}, q \in [0, \pi]  \nn 
\eea
where $J\equiv J_1 = J_2 = K/4$. The spin-triplet excitation spectrum of 
an SU(4) chain with $N= 4 n$ sites, shown in Fig. 3(a), contains convolutions 
of elementary $\sigma$ and $\tau$ excitations, as well as $\nu$ excitations. 
The $z$-component of the dynamical spin structure factor $S^z(k,\omega)$ 
couples directly to the 15-fold multiplet 
of $\sigma - \tau$ pairs, the 20-fold multiplet of $2 \nu$
excitations, and with much smaller matrix elements to the 
45-fold $2 \sigma - \nu$ and the 35-fold $4 \sigma$ multiplets.\cite{li}

Fig. 3(b) illustrates that the spectrum obtained with the dynamical
DMRG traces exactly the two-soliton continua, spanned by 
$\omega_{\sigma}(q) + \omega_{\tau} (q')$ and by $\omega_{\nu} (q) 
+ \omega_{\nu} (q')$. These continua have characteristic low-frequency
power-law singularities ($S^z(k,\omega) \propto \omega^{-\alpha}$) at
the Fermi vectors $\pi/2$ and $3 \pi/2$, and a weaker divergence at
$\pi$.\cite{footnote} The relatively high spectral intensity seen at 
larger frequencies and centered around the wave vector $\pi$, can be 
attributed to a particle-hole excitation in the $\nu$-channel which 
will be discussed later. We note that the accuracy of the dynamical 
DMRG method is known to deteriorate at larger energies, and hence the 
spectral intensities are less reliable in this regime.

When an external magnetic field is applied along the $z$-direction,
the spectrum of the SU(4) chain
also becomes incommensurate.\cite{yamashita2} This effect
can be understood quantitatively by considering the splitting of spin-up and
spin-down bands due to the applied field. The corresponding 
nesting vectors depend on the magnetization $m \in [0, 1/2]$ as
$2 k_{F \uparrow} = \pi ( 1 + 2 m)/2$, 
$2 k_{F \downarrow} = \pi ( 1 - 2 m)/2$, and
$4 k_{F \uparrow } = 4 k_{F \downarrow } =
\pi ( 1 - 2 m)$. This low-frequency behavior
is reflected in the finite-size DMRG spectra, shown in Fig. 4, with
magnetizations $m=4/40$ and $m=12/40$. For example, with $m=4/40$, the 
nesting vectors are $6\pi /10$, $4\pi /10$, and $8\pi /10$, as shown in 
Figs. 4(a) and (c), whereas for $m=12/40$ they are $8\pi /10$, $2\pi /10$, and 
$4\pi /10$ respectively (Figs. 4(b) and (d)).

Furthermore, the overall spectral weight is reduced as the phase space 
of spin-triplet excitations shrinks with increasing field. At finite 
magnetic fields, there remains a commensurate soft mode at wave vector 
$\pi$ in the orbital $\nu$-channel which does not couple to $S(k,\omega)$. 
It is therefore interesting to examine the orbital excitation spectrum 
traced out by $T(k,\omega)$. 
The orbital dynamical structure factor of the SU(4) chain in a magnetic
field is shown in Figs. 4 (c) and (d). In the absence of a magnetic
field, the orbital spectrum is identical to the spin spectrum 
(Fig. 3). In a finite magnetic field
there is one orbital mode, the spin-1 $\nu$-channel,
which does not become incommensurate, but stays soft at wave vector
$\pi$ for all magnetic fields.  On the other hand, the strongly 
mixed $\sigma - \tau$ pairs show incommensuration with the same nesting 
vectors as for the spin channel. The dominant
$\sigma - \tau$ singularities at low frequencies occur
at $2 k_{F \uparrow}$ for the orbitons and at $2 k_{F \downarrow}$
for the spinons.\cite{yamashita2} 

The excitation spectra at the Majumdar-Ghosh point ($J_1=J_2=3K/4$) are 
shown in Fig. 5. The zero-field ground state at this 
point in the dimerized phase IV is a doubly degenerate product 
of spin and orbital singlets with an energy of $-3J_1 /4$ per site and
a gap of approximately $3K/8$ to the lowest excitations.\cite{pati,mg,kolezhuk}
The spectrum contains a 2-soliton 
continuum of states with minima at wave vectors $k = 0$ and $\pi$, 
and maximum spectral weight at wave vector $\pi$.\cite{kolezhuk}
In addition, there are massive magnon excitations corresponding 
to Haldane triplet bound states.\cite{haldane} These are centered at 
$k = \pi /2$ and $3\pi /2$, indicating a doubling of the 
real-space unit cell. In Fig. 5(a), we have eliminated the
artificial low-energy states which arise from surface excitations
at the open boundaries of the 80-site cluster. These states 
have vanishing spectral weight in the thermodynamic limit.

A finite magnetic field leads to the deconfinement of bound 
spinons.\cite{yu} This is manifested in a lowering and eventual 
vanishing of the corresponding spin gap when the field exceeds a 
critical value, which for the Majumdar-Ghosh point is at $h_{c1} 
\approx 3K/8$. In Fig. 5(b), the spin
excitation spectrum is shown in this incommensurate regime ($h_{c1}
< h < h_{c2}$) at a magnetization $ m = 4/40$. The dominant low-energy
singularities of the spin-1 magnons occur at wave vectors  
$4 k_{F} = \pi ( 1 \pm 2 m)$. The 2-spinon continuum at higher energies
appears to be almost unaffected by the finite magnetic field.
Interestingly, a strong low-energy singularity at $k = \pi$
in the orbital channel is induced by the applied field, as 
observed in Fig. 5(c), which is indicative of an effective attractive 
potential between the incommensurate spin-1 and the commensurate
orbital-1 $\nu$-type solitons in this case. This magnetic-field-induced
two-component Luttinger Liquid behavior,
with the orbital channel remaining commensurate and the spin channel becoming 
incommensurate, has also been observed in recent bosonization studies of the
spin-orbital chain.\cite{orignac,lee} 

Finally, let us turn to the gapless phase V.
In Figs. 6 and 7, the spin and orbital excitation spectra are shown at
$J_1=J_2=K/8$ and at $J_1=J_2=0$. As $K \rightarrow 0$, a prominent
high-energy excitation emerges, centered at wave vector $\pi$
and frequency $\omega \approx 0.8K$.
This excitation carries the quantum numbers $(S,T)$ = (0,1) and (1,0)
which indicates a particle-hole-type bound state, involving a spin singlet
and an orbital triplet or {\it vice versa}. From finite-size scaling, we 
observe that the energy gap between the ground state and this level
increases with system size. If $J_1=J_2=0$, the eigenstates of the
$K$ term can be found exactly on a 4-site cluster. The ground state with
energy $-3K/4$ and the excitons with energy 0 are in this case
\bea
|GS \rangle &=& \sum_{i=1}^4  (-1)^i
| (t_{i,O}^+ t_{i+2,O}^- ) (s_{i,S} s_{i+2,S} ) -
(s_{i,O} s_{i+2,O} ) (t_{i,S}^+ t_{i+2,S}^- ) \nn\\
 &+&
(t_{i+1,O}^- t_{i+3,O}^+ ) (t_{i,S}^+ t_{i+2,S}^- ) -
(t_{i,O}^+ t_{i+2,O}^- ) (t_{i+1,S}^- t_{i+3,S}^+ )\rangle  \nn\\
|\pi, S \rangle &=& \sum_{i=1}^4
| (t_{i,O}^+ t_{i+2,O}^- ) (t_{i,S}^+ s_{i+2,S} ) \rangle \ \ {\rm and} \ \ 
|\pi, O \rangle = \sum_{i=1}^4
| (t_{i,S}^+ t_{i+2,S}^- ) (t_{i,O}^+ s_{i+2,O} ) \rangle \nn
\eea
where, for example, $t_{i,S}^+$ denotes an up-up triplet state
in the spin channel, involving
spins at sites $i$ and $i+1$, and $s_{i+2,O}$ denotes a singlet in the 
orbital channel at sites $i+2$ and $i+3$. This high-energy bound state 
is observed in the entire phase V, but is most prominent in the 
strong-$K$ limit.   

The magnetic-field dependence of the spectra in phase V can be 
analyzed analogously to the SU(4) limit. In particular, the
zero-energy modes can be understood within the band picture
discussed above, with nesting instabilities at $2k_{F \uparrow}$, 
$2k_{F \downarrow}$, and $4k_{F \uparrow}$=$4k_{F \downarrow}$. 
The finite-size gaps in these spectra appear to be
most obvious at the critical point, and become
less severe as $J_1,J_2 \rightarrow 0$. 

\section{Summary}

In summary, the spin and orbital excitation spectra of the 
spin-orbital chain were studied in various regimes of the model 
Hamiltonian $H$, given in Eq. 1. The characteristic dynamical 
response depends strongly on the parameter set ($J_1,J_2,K$) and 
on the applied field $h$. In phases II and III
with either fully polarized spin or orbital degrees of freedom,
the remaining gapless mode has the characteristic spectral
response of a (pseudo)-spin-1/2 Heisenberg chain. On the other
hand, in the dimerized regime (phase IV), a finite magnetic field 
is needed to close the spin gap. Interestingly, an additional 
orbital mode is found to become gapless at the commensurate wave 
vector $\pi$ in the partially spin-polarized regime ($h_{c1} < h 
< h_{c2}$), whereas the Fermi momenta of the incommensurate
gapless spin modes depend on the magnetization via
$2k_F = \pi (1 - 2 m)$. In the gapless regime (phase V), 
there are three elementary gapless excitations.
The dynamical response consists of various overlapping
two-particle continua, arising from pairs of these elementary 
excitations. At higher energies, a particle-hole bound state 
whose origin lies in the biquadratic $K$-term, is found to 
dominate the spectrum.

We are greatful to Andreas Honecker, Till K\"uhner, Bruce Normand,
and Edmond Orignac
for useful discussions, and we thank Till K\"uhner for valuable
help in developing the dynamical DMRG algorithm.
S. H. acknowledges the Zumberge foundation for financial support.

\newpage

\begin{figure}[h]
\centerline{\psfig{figure=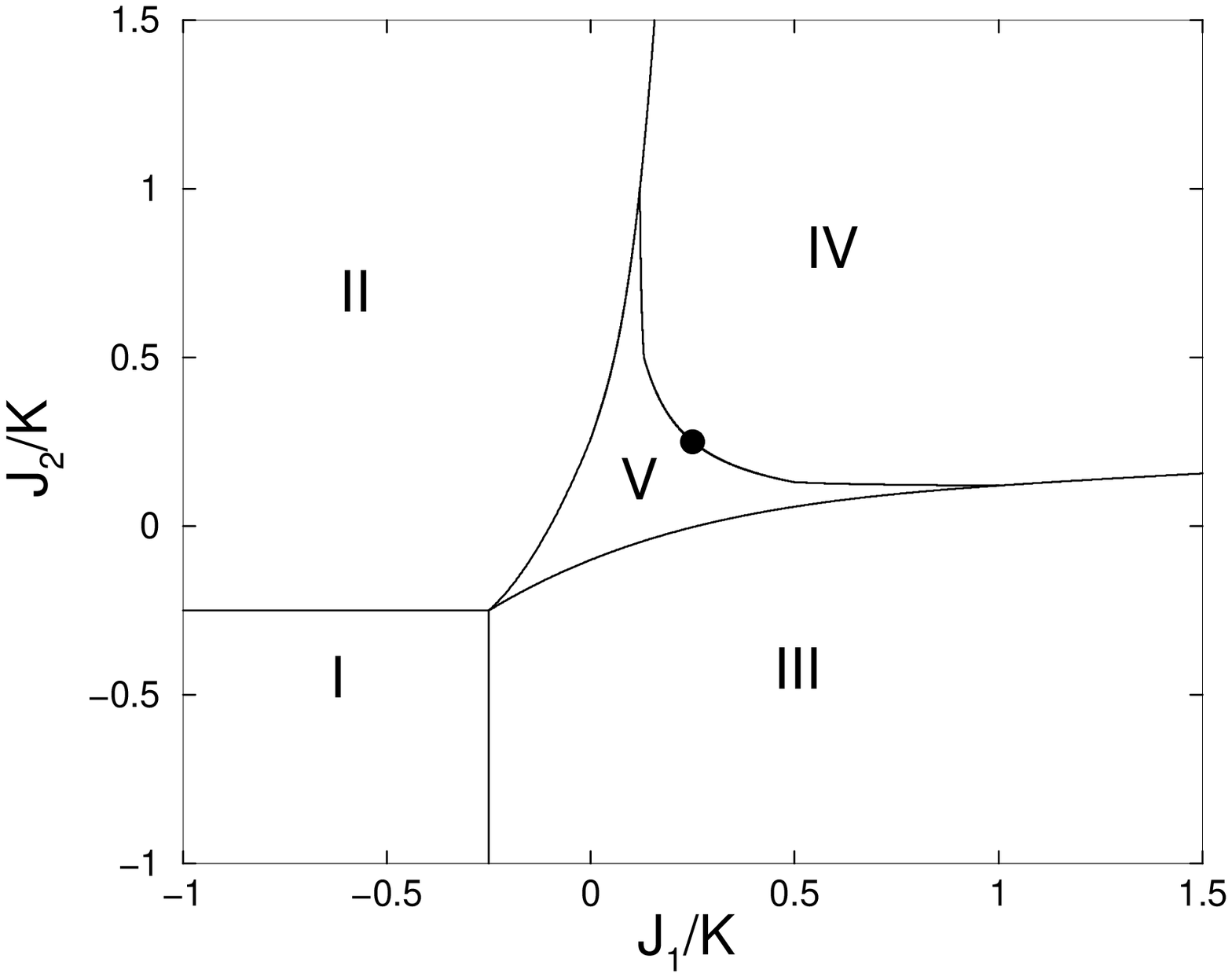,width=11cm}}
\vspace{0.3cm}
\caption{
Phase diagram of the one-dimensional spin-orbital model, adapted from
Ref. 3. Phase I: spins and orbitals are fully polarized. 
Phases II and III: one of the channels is fully polarized, whereas the
other is a quasi-long-range-ordered antiferromagnet. Phase IV:
dimerized regime. Phase V: gapless regime. Circle: SU(4)-symmetric point.
}
\end{figure}

\begin{figure}[h]
\centerline{\psfig{figure=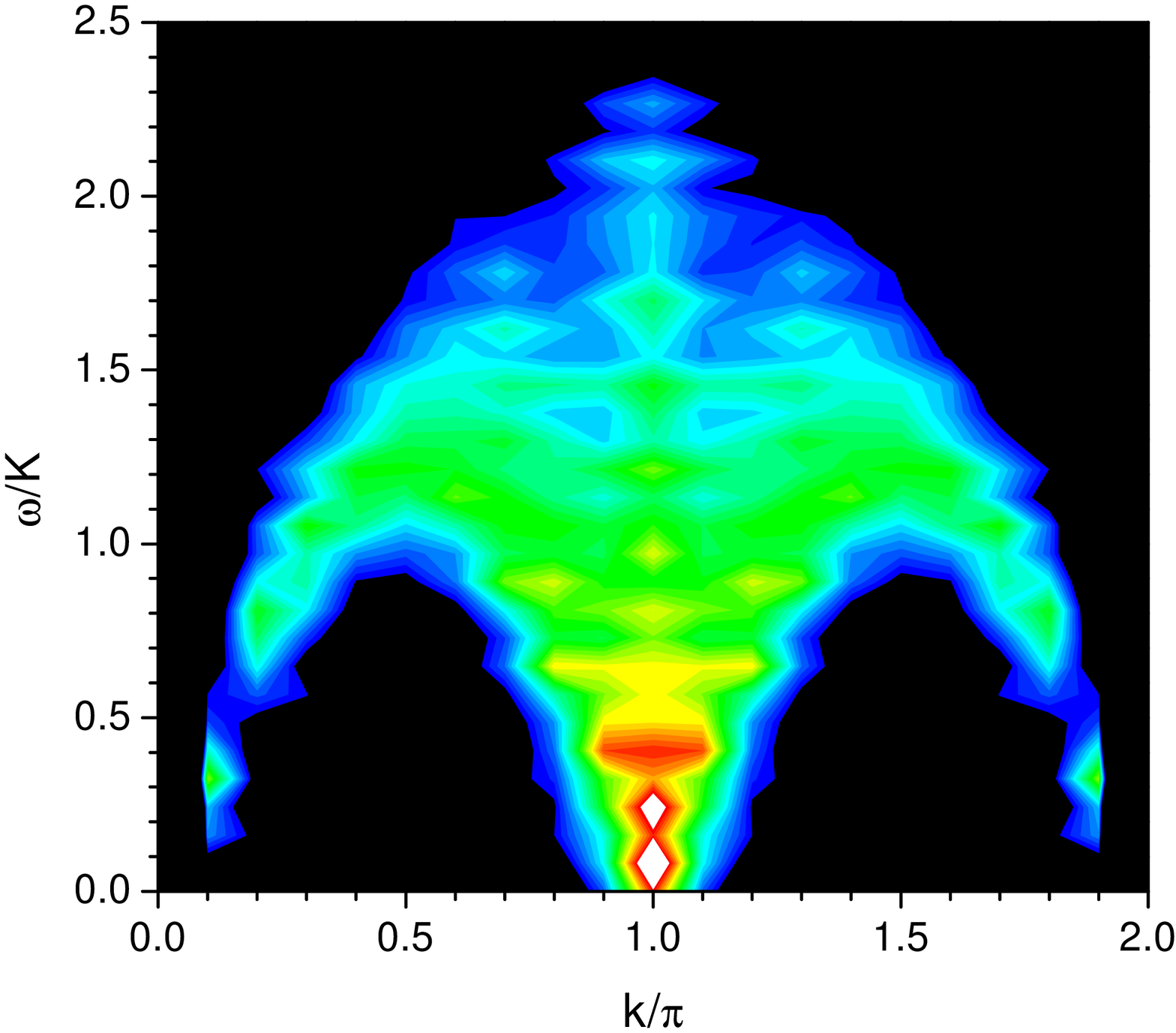,height=8cm}
\psfig{figure=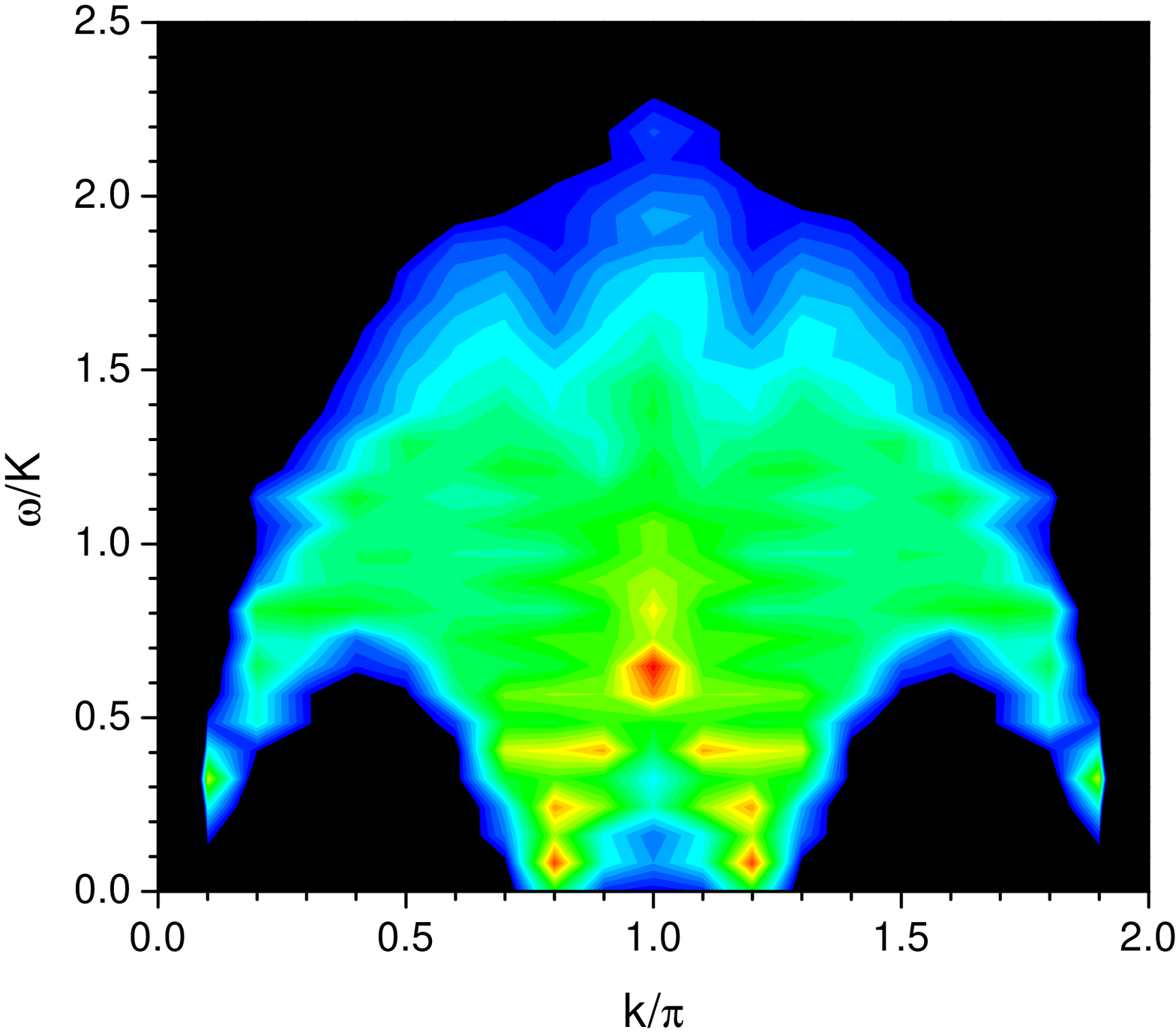,height=8cm}}
\vspace{0.3cm}
\caption{
Spin excitation spectrum of the spin-orbital chain at ($J_1=-J_2=K/2$), 
obtained with dynamical DMRG on a 80-site cluster. (a) zero magnetic
field $m = 0$, and (b) $m$ = 4/40.
}  
\end{figure}

\begin{figure}[h]
\centerline{\psfig{figure=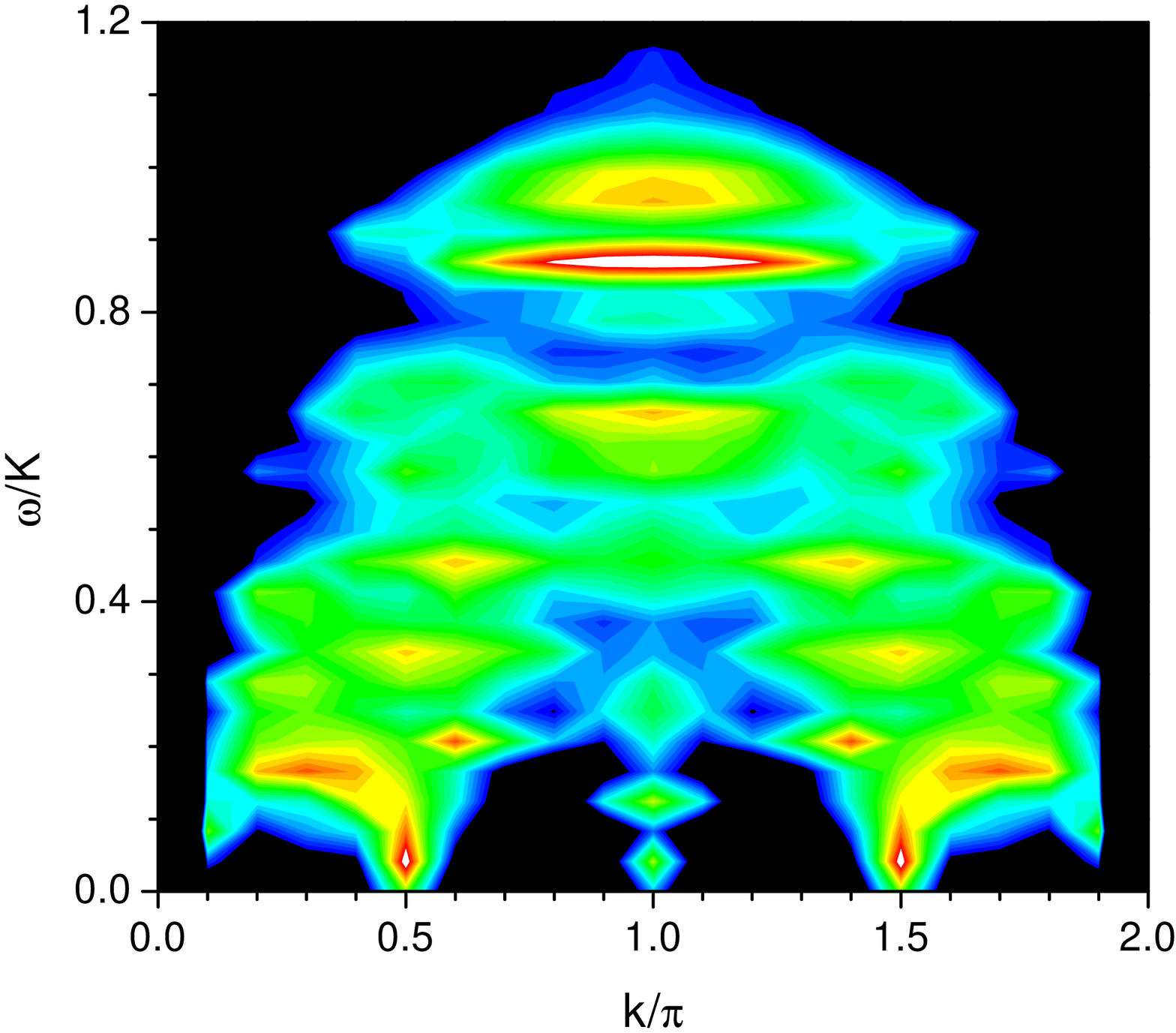,height=8cm}
\psfig{figure=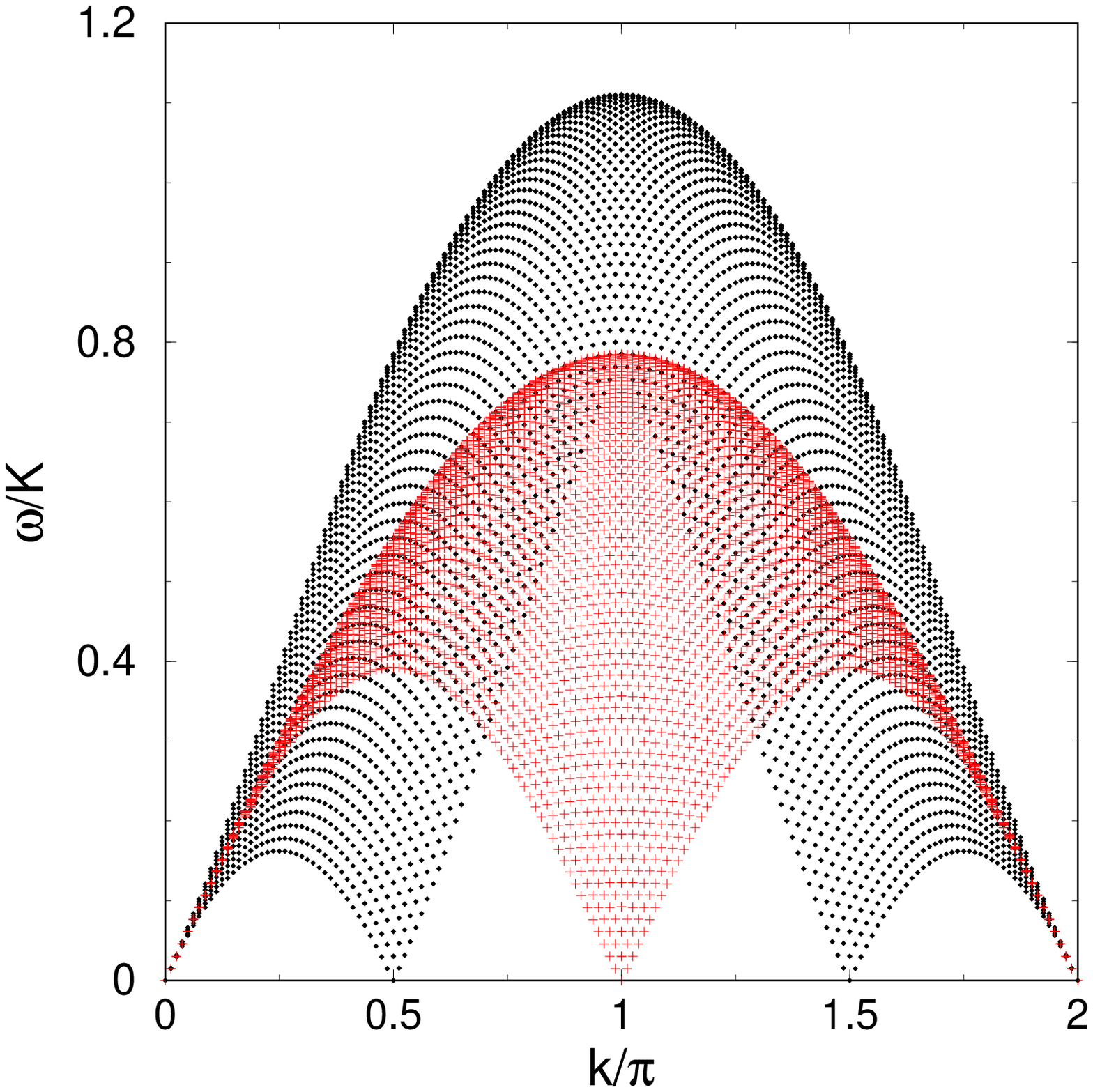,angle=270,height=8cm}}
\vspace{0.3cm}
\caption{
(a) Dynamical spin structure factor of the SU(4)-symmetric
spin-orbital chain. This spectrum was calculated using the dynamical
DMRG technique on an 80-site cluster.
(b) Spin-triplet spectra composed of $\sigma - \tau$ pairs (dots) 
and $2 \nu$ pairs (pluses), obtained from convolutions of the
single-soliton dispersions, which are known exactly from the 
Bethe Ansatz solution.
}
\end{figure}

\begin{figure}[h]
\centerline{\psfig{figure=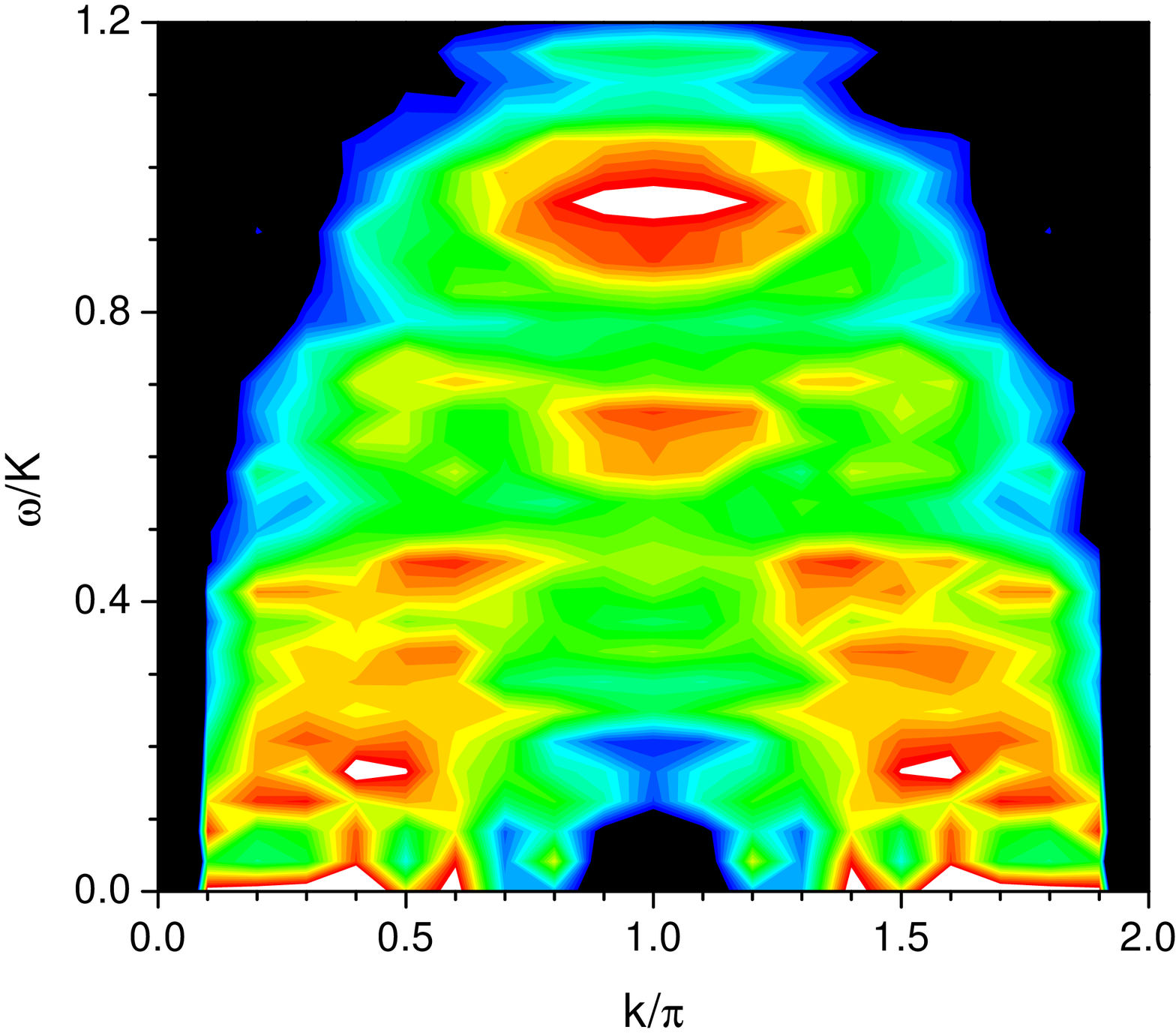,height=8cm}\psfig{figure=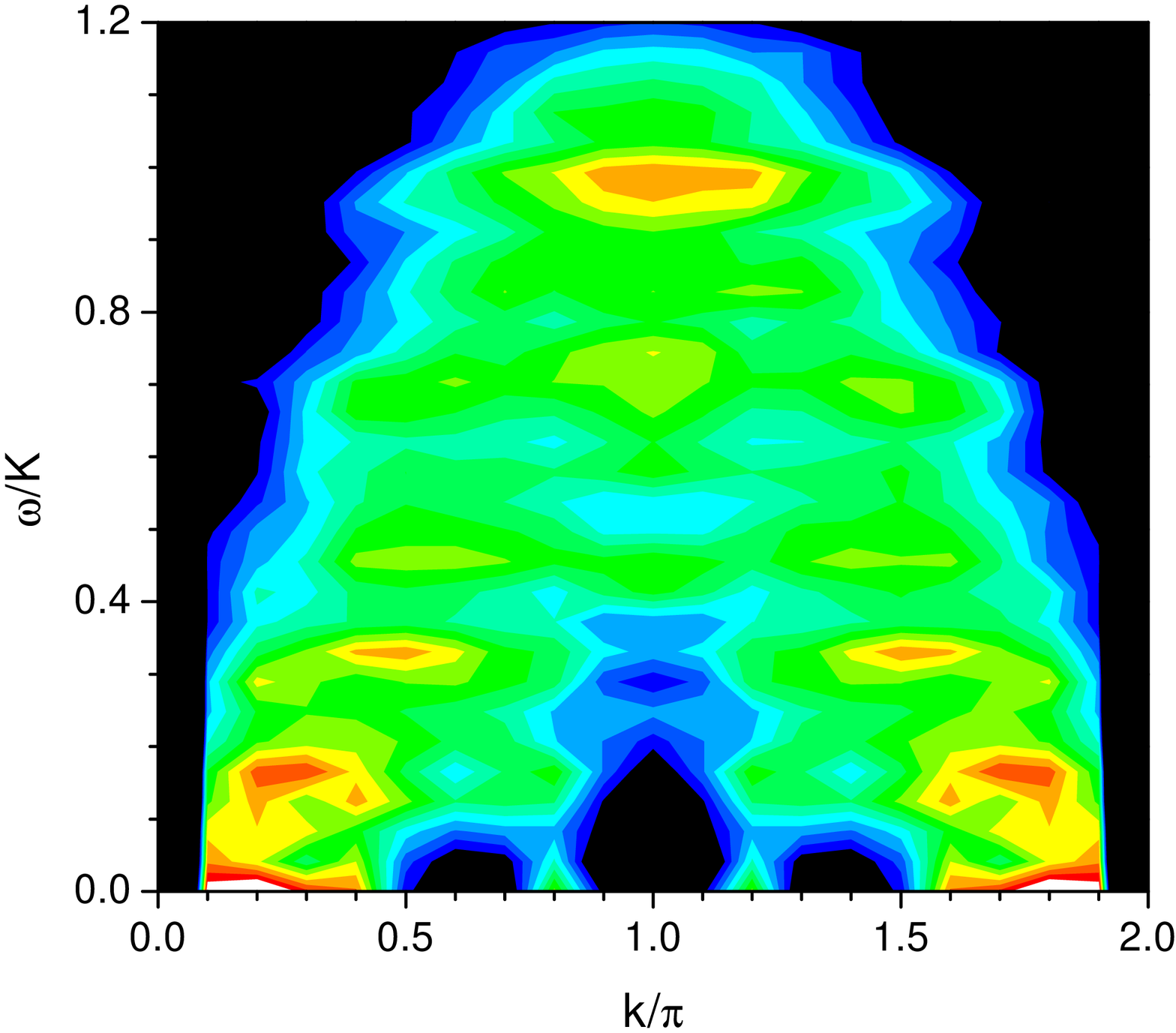,height=8cm}}
\centerline{\psfig{figure=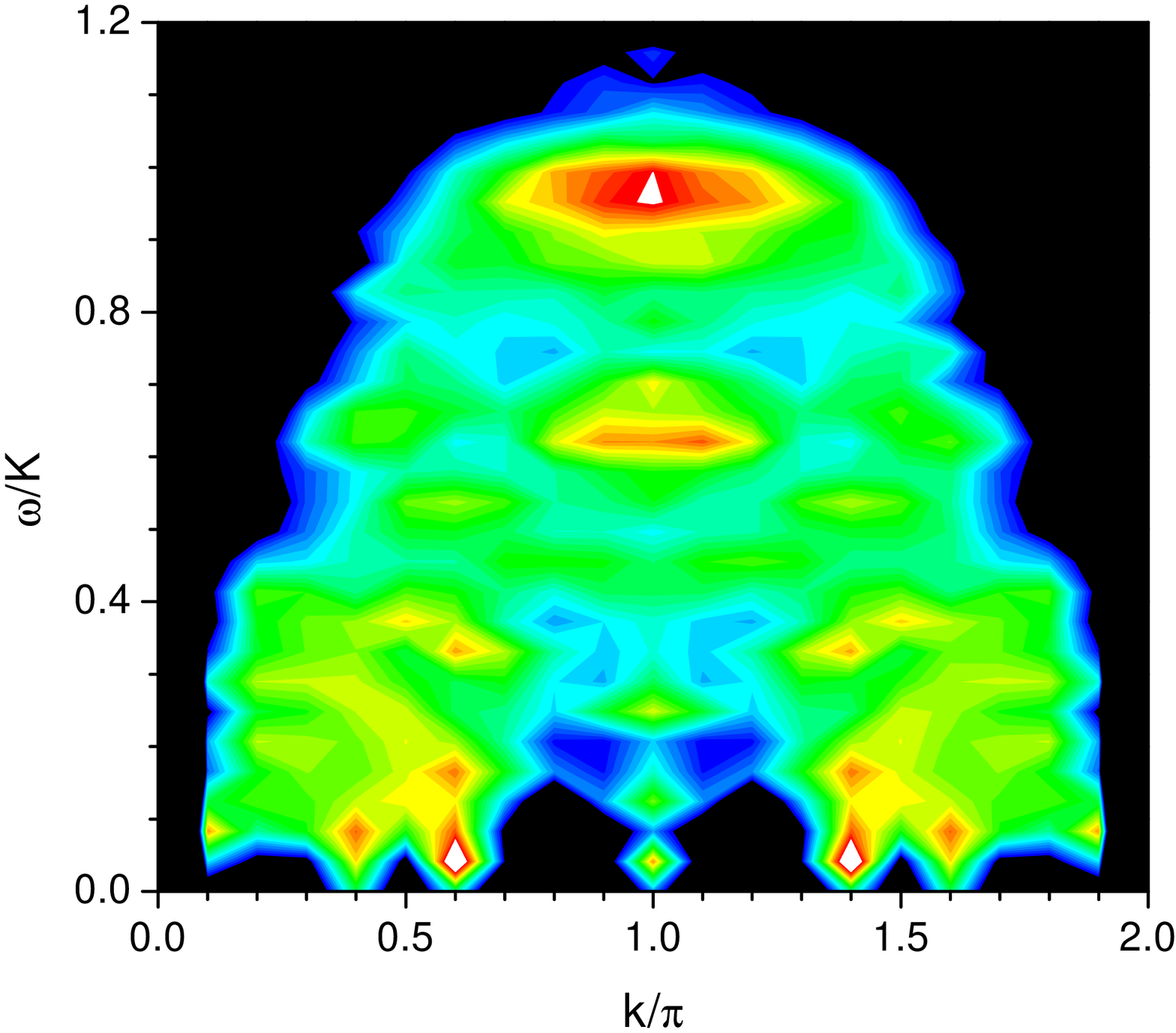,height=8cm}\psfig{figure=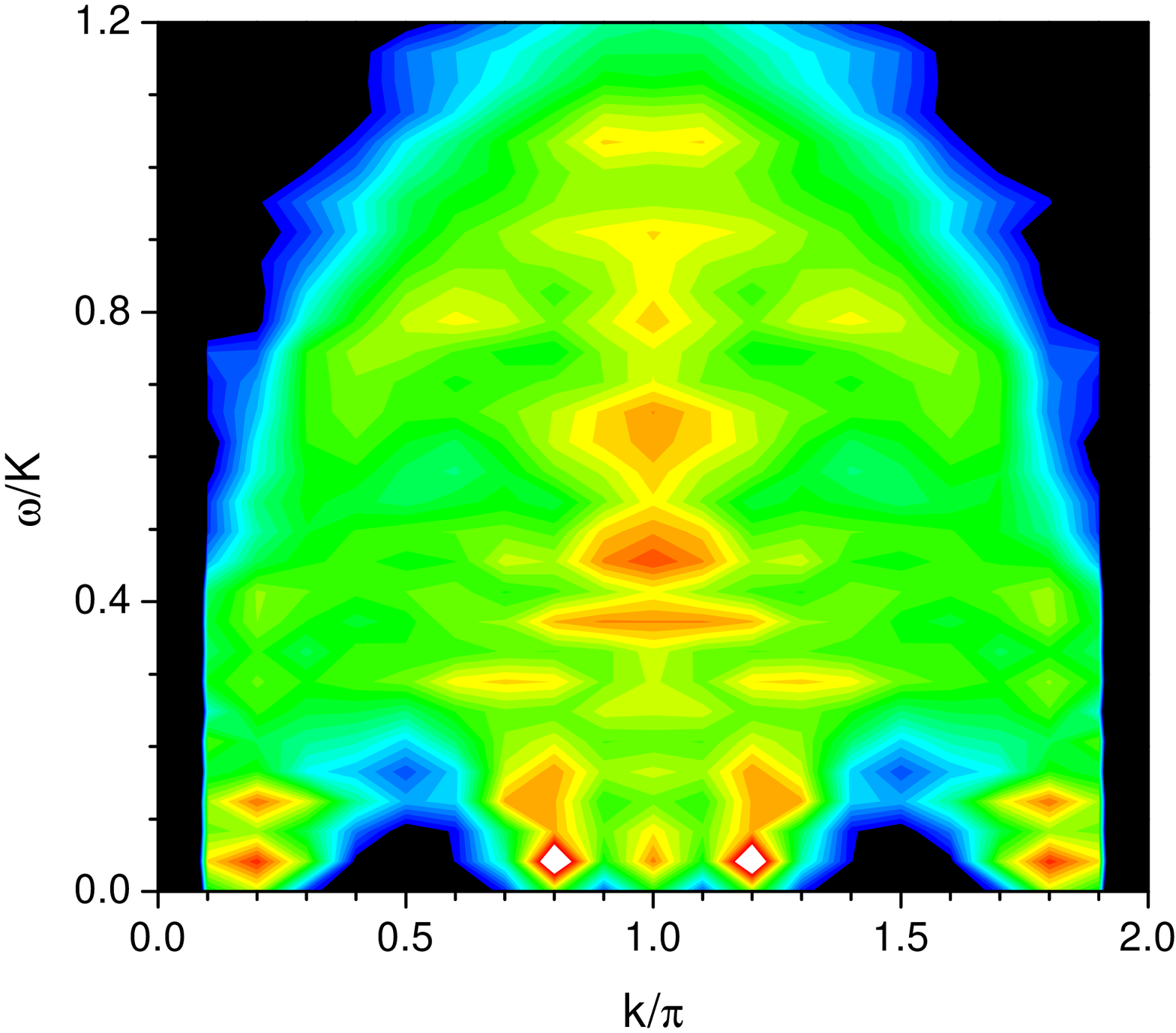,height=8cm}}
\vspace{0.3cm}
\caption{Spin and orbital excitation spectra of the SU(4) chain in a 
magnetic field. (a) (upper left) spin dynamics at $m=4/40$, 
(b) (upper right) spin dynamics
at $m=12/40$, (c) (lower left) orbital dynamics at $m=4/40$, and (d) 
(lower right) orbital dynamics at $m=12/40$.
}
\end{figure}

\begin{figure}[h]
\centerline{\psfig{figure=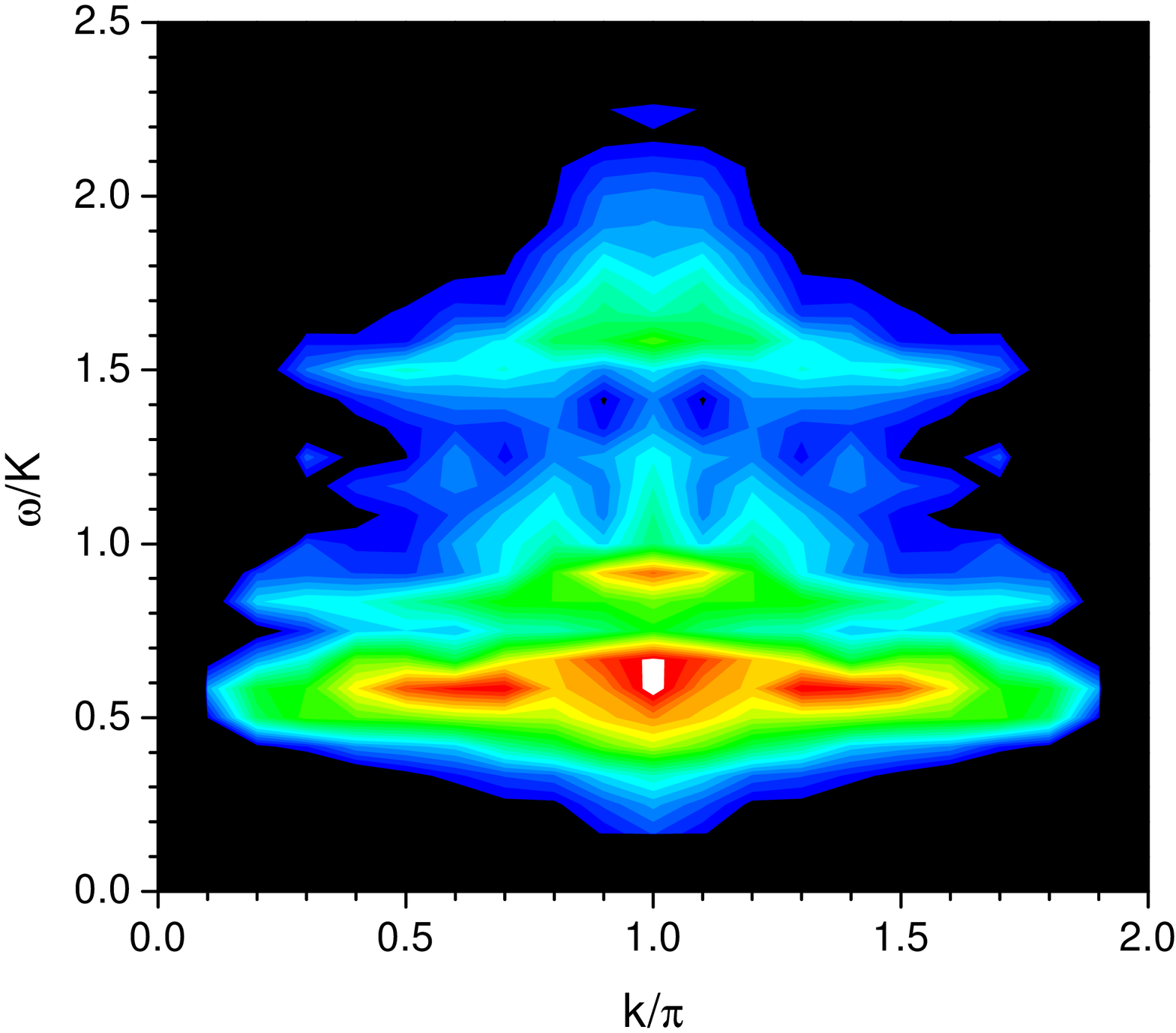,height=8cm}}
\centerline{\psfig{figure=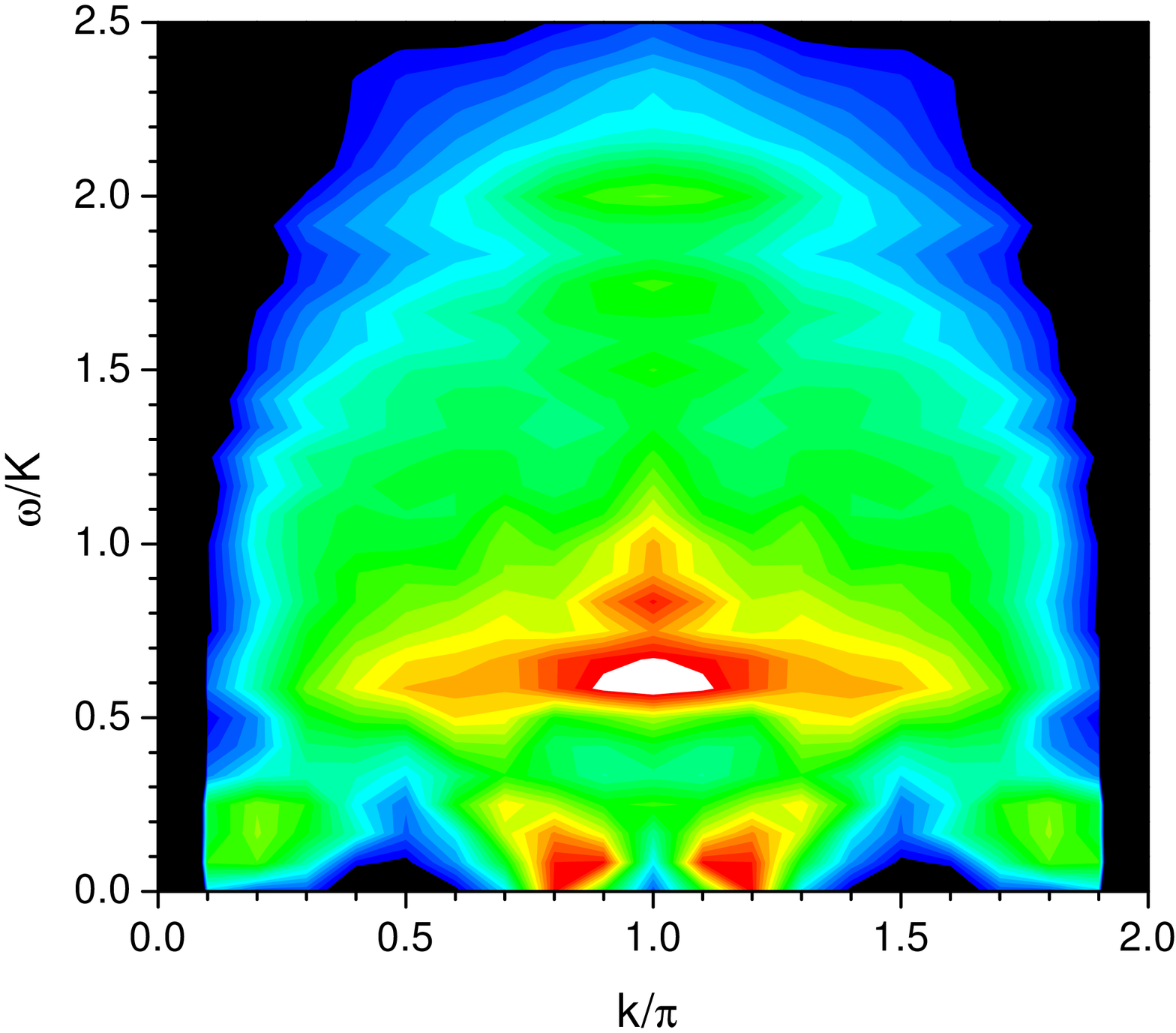,height=8cm}
\psfig{figure=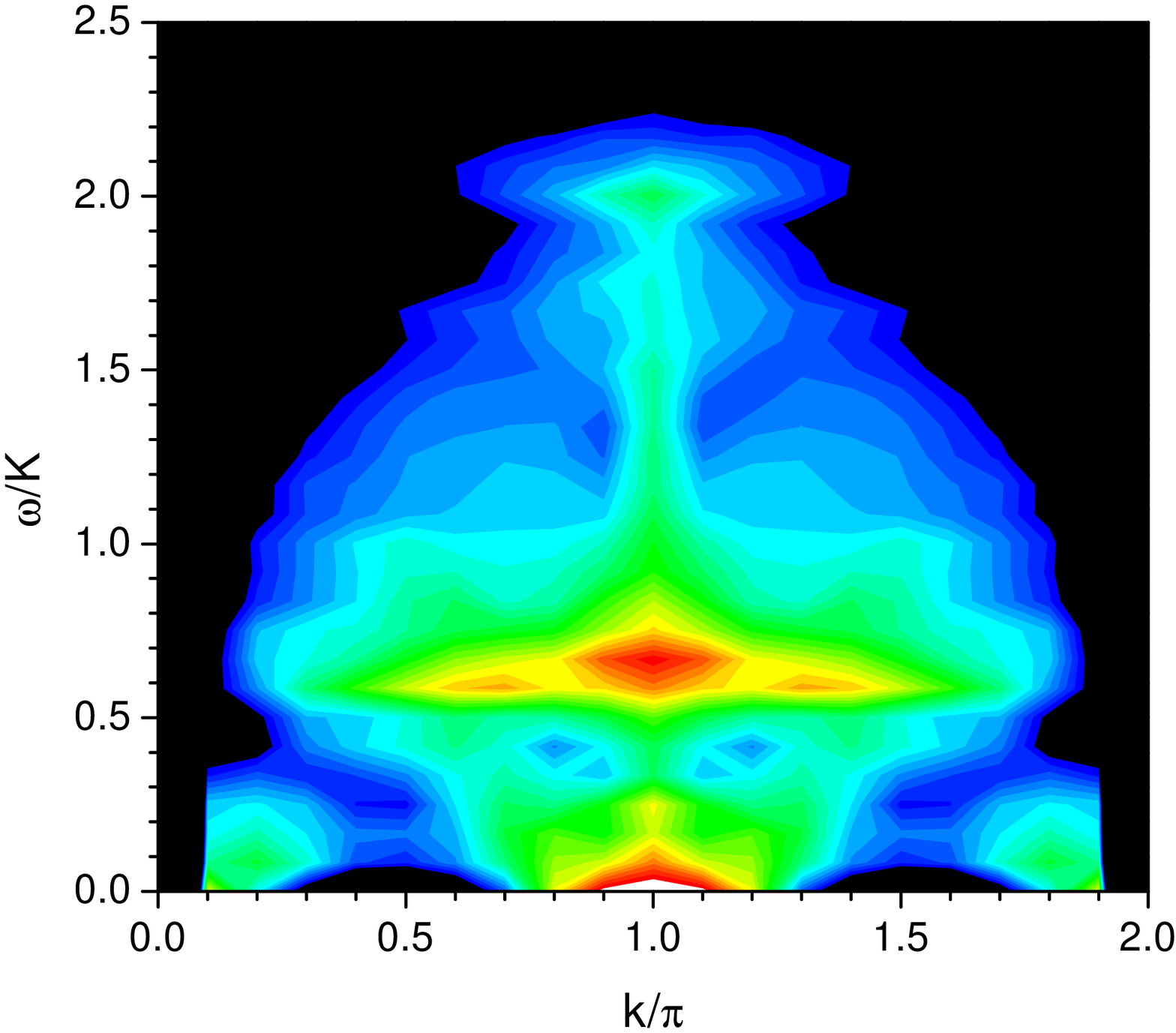,height=8cm}
}
\vspace{0.3cm}
\caption{Spin and orbital excitation spectra of the spin-orbital chain
at the Majumdar-Ghosh point ($J_1=J_2=3K/4$). 
(a) (upper) spin and orbital dynamics at $m=0$,
(b) (lower left) spin dynamics at $m=4/40$, and (c)
(lower right) orbital dynamics at $m=4/40$.
}
\end{figure}

\begin{figure}[h]
\centerline{\psfig{figure=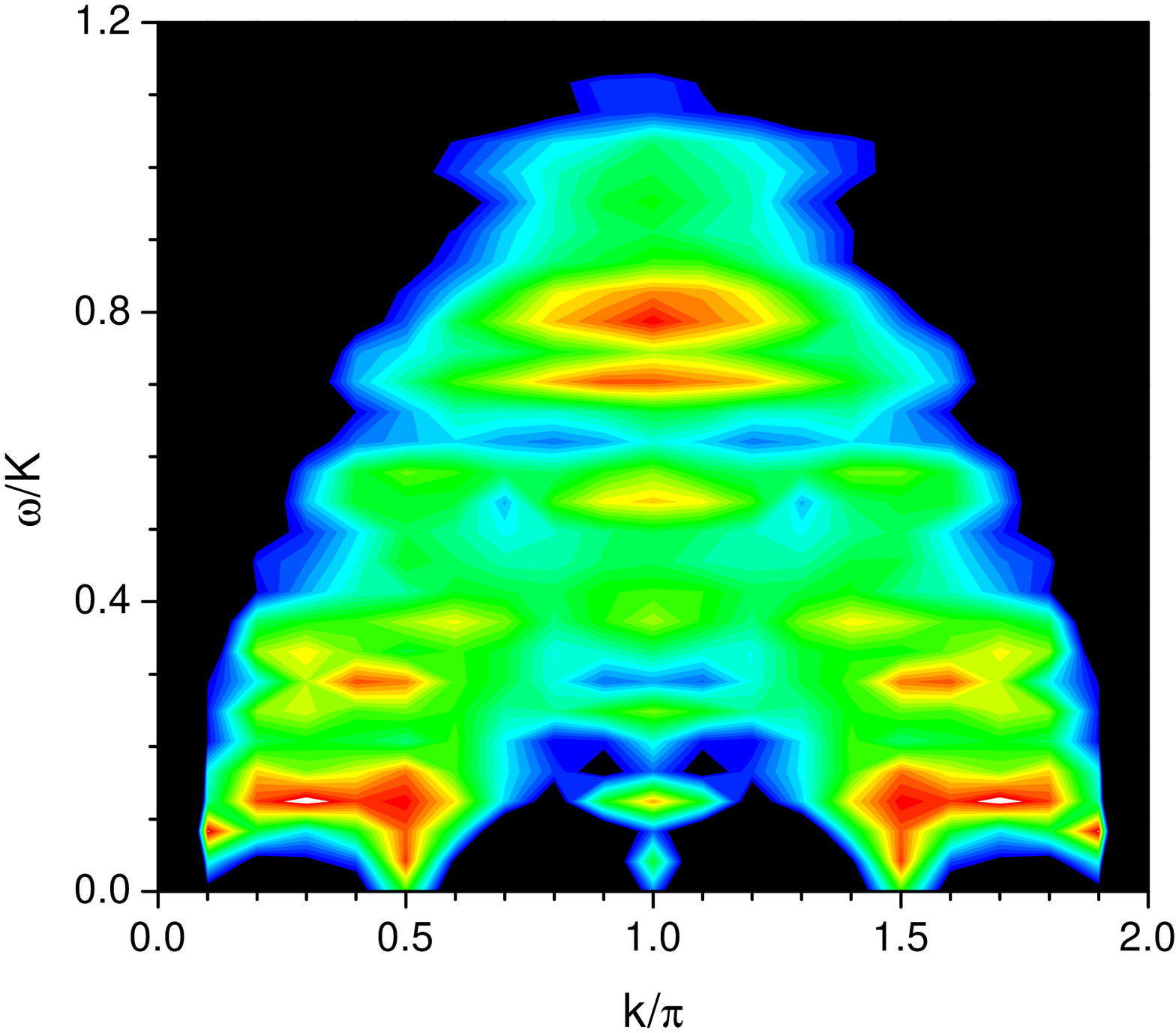,height=8cm}}
\centerline{\psfig{figure=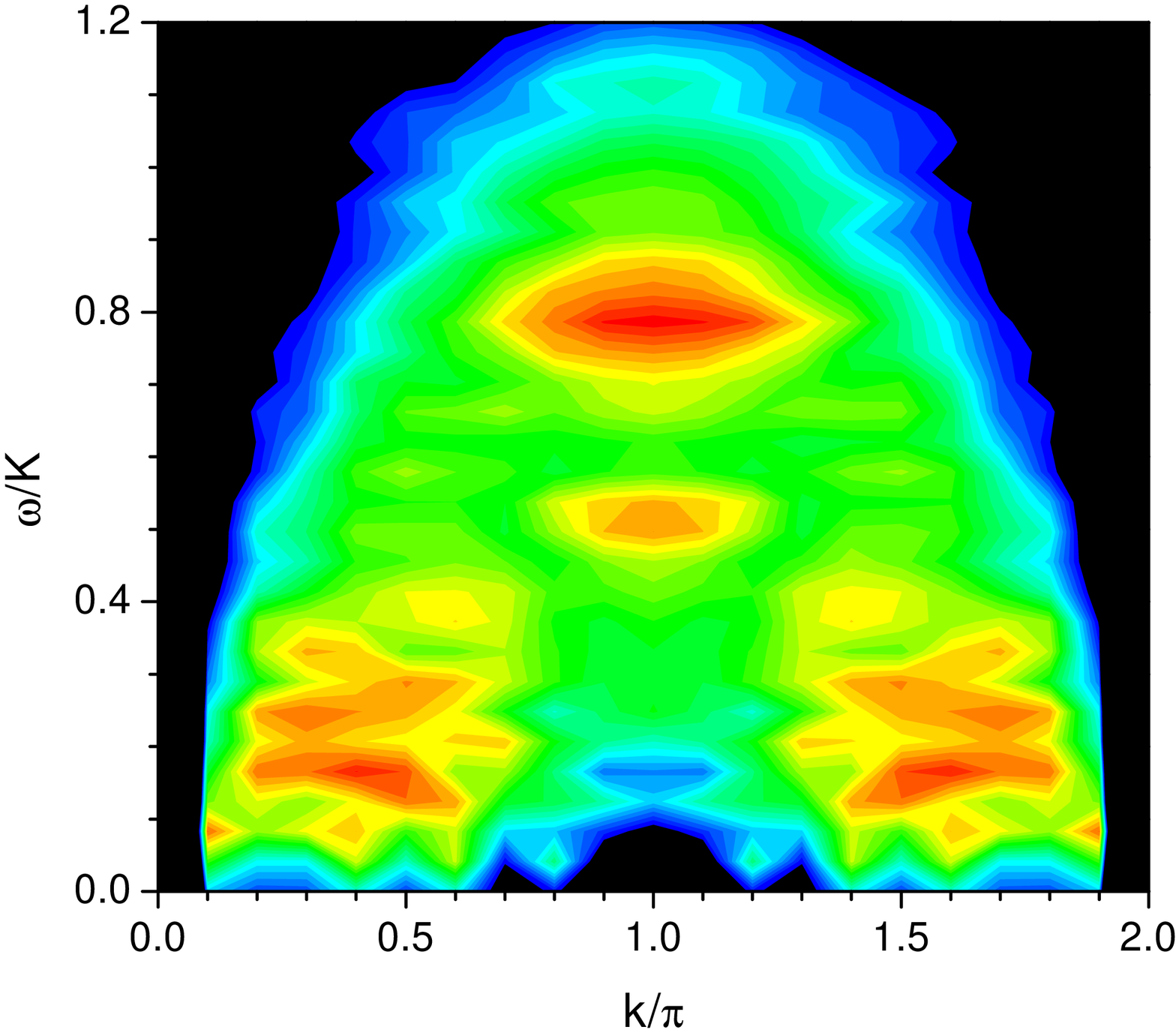,height=8cm}
\psfig{figure=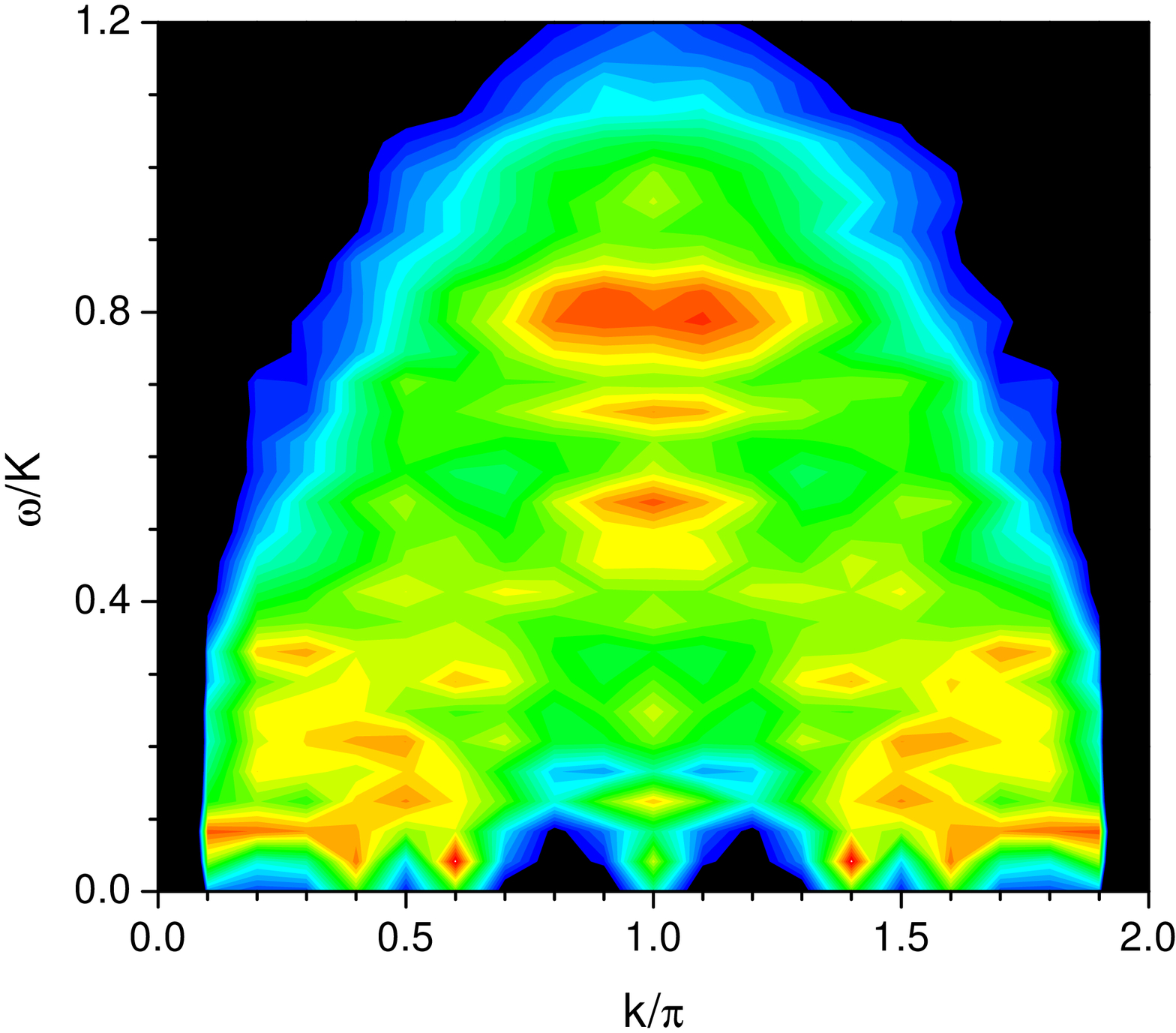,height=8cm}
}
\vspace{0.3cm}
\caption{Spin and orbital excitation spectra of the spin-orbital chain
at ($J_1=J_2=K/8$).
(a) (upper) spin and orbital dynamics at $m=0$,
(b) (lower left) spin dynamics at $m=4/40$, and (c)
(lower right) orbital dynamics at $m=4/40$.
}
\end{figure}

\begin{figure}[h]
\centerline{\psfig{figure=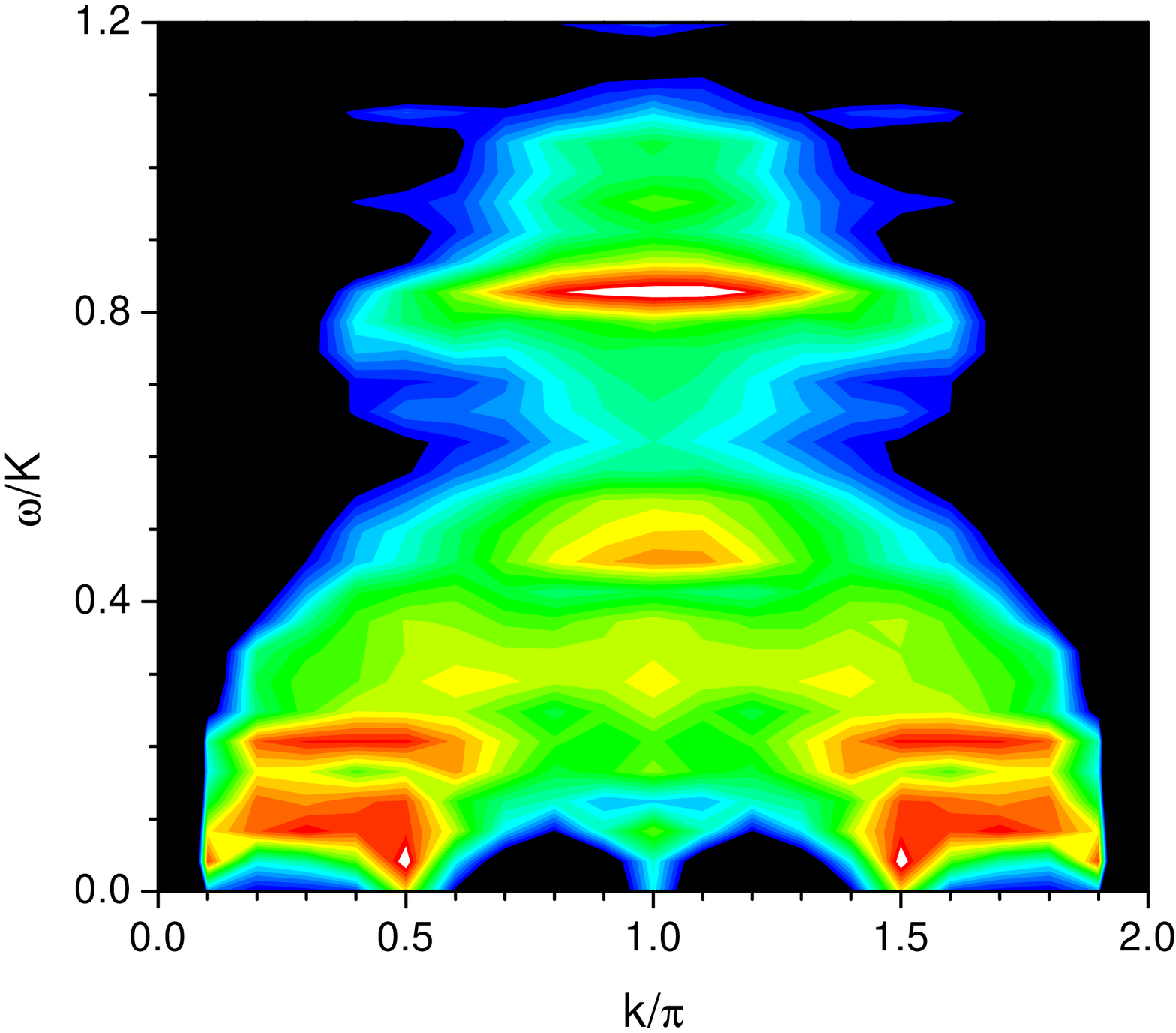,height=8cm}}
\centerline{\psfig{figure=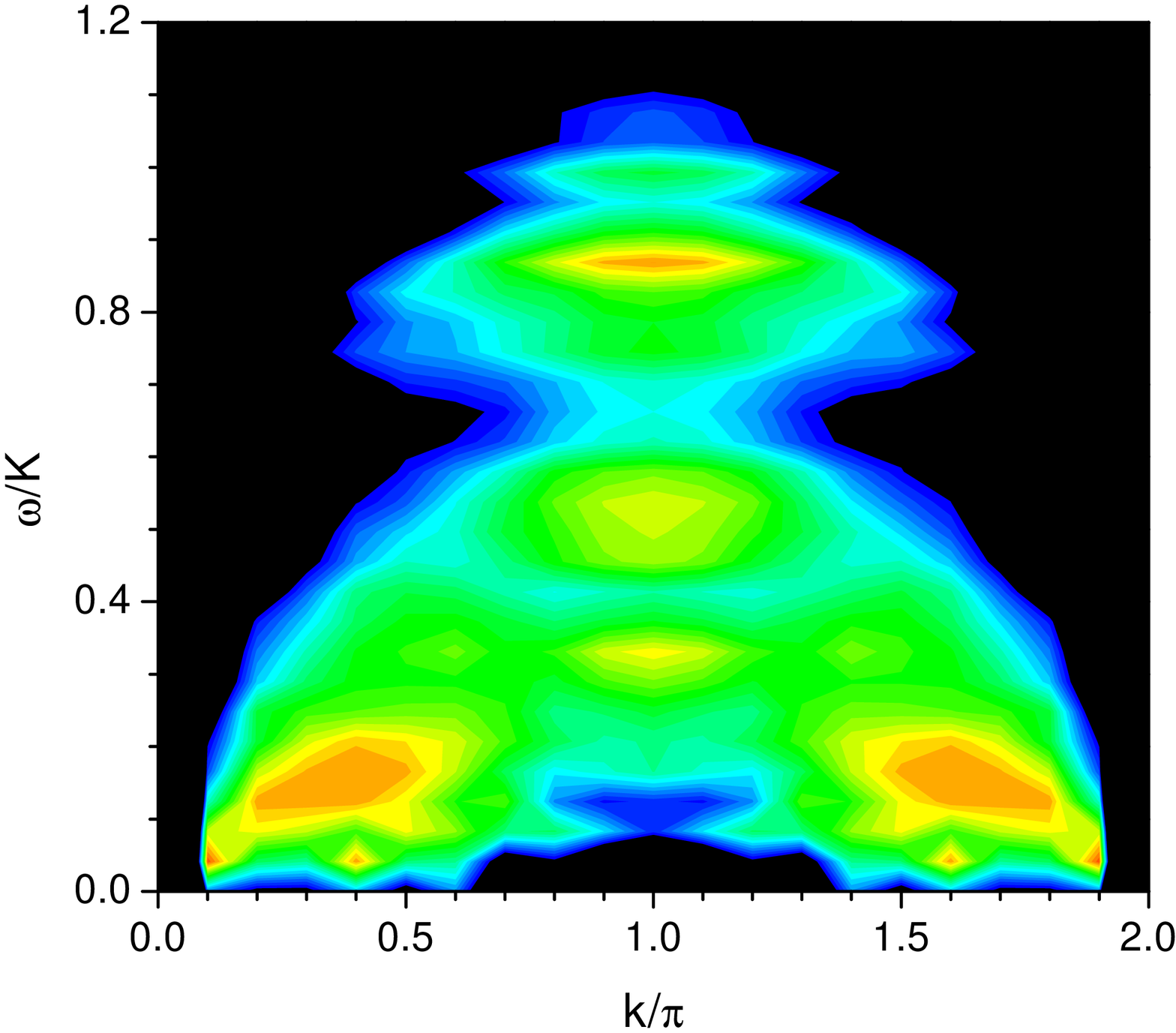,height=8cm}
\psfig{figure=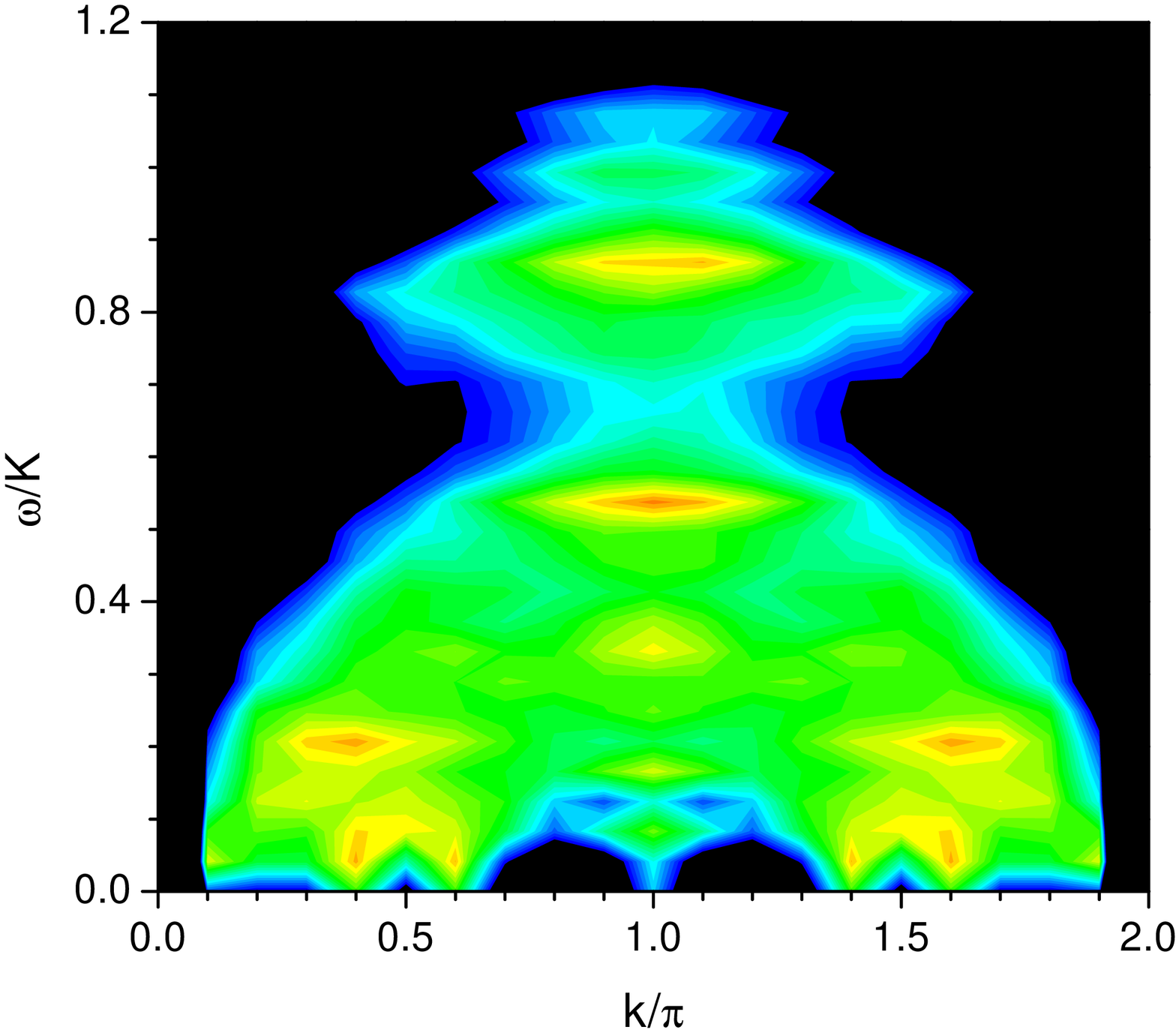,height=8cm}
}
\vspace{0.3cm}
\caption{Spin and orbital excitation spectra of the spin-orbital chain
at ($J_1=J_2=0$).
(a) (upper) spin and orbital dynamics at $m=0$,
(b) (lower left) spin dynamics at $m=4/40$, and (c)
(lower right) orbital dynamics at $m=4/40$.
}
\end{figure}


\begin{references}
\bibitem{footnote} Logarithmic corrections to these low-energy singularities
are expected to occur. However, a comprehensive analysis of the scaling 
behavior, including the effects of
marginally relevant currents, is beyond the scope of
this work, and will be addressed in a future publication.
\end{references}
\end{document}